\documentclass[sigconf]{acmart}
\usepackage{multirow}
\usepackage{booktabs}
\usepackage{enumitem}
\usepackage[linesnumbered,ruled,vlined]{algorithm2e} 
\usepackage{algpseudocode}
\usepackage{amsmath}  
\usepackage{balance}

\AtBeginDocument{%
  \providecommand\BibTeX{{%
    \normalfont B\kern-0.5em{\scshape i\kern-0.25em b}\kern-0.8em\TeX}}}

\copyrightyear{2020}
\acmYear{2020}
\setcopyright{acmcopyright}
\acmConference[CIKM '20]{The 29th ACM International Conference on Information and Knowledge Management}{October 19--23, 2020}{Virtual Event, Ireland}
\acmBooktitle{The 29th ACM International Conference on Information and Knowledge Management (CIKM '20), October 19--23, 2020, Virtual Event, Ireland}
\acmPrice{15.00}
\acmDOI{10.1145/3340531.3412086}
\acmISBN{978-1-4503-6859-9/20/10}

\begin{document}
\fancyhead{}
\title{Data Augmentation for Graph Classification}


\author{Jiajun Zhou}
\affiliation{%
  \institution{Zhejiang University of Technology}
  \city{HangZhou}
  \country{China}}
\email{jjzhou@zjut.edu.cn}
\author{Jie Shen}
\affiliation{%
  \institution{Zhejiang University of Technology}
  \city{HangZhou}
  \country{China}}
\email{shenj@zjut.edu.cn}
\author{Qi Xuan}
\affiliation{%
  \institution{Zhejiang University of Technology}
  \city{HangZhou}
  \country{China}}
\email{xuanqi@zjut.edu.cn}
\begin{abstract}
Graph classification, which aims to identify the category labels of graphs, plays a significant role in drug classification, toxicity detection, protein analysis etc.
However, the limitation of scale of benchmark datasets makes it easy for graph classification models to fall into over-fitting and undergeneralization.
Towards this, we introduce data augmentation on graphs and present two heuristic algorithms: {\em{random mapping}} and {\em{motif-similarity mapping}}, to generate more weakly labeled data for small-scale benchmark datasets via heuristic modification of graph structures. Furthermore, we propose a generic model evolution framework, named {\em{M-Evolve}}, which combines graph augmentation, data filtration and model retraining to optimize pre-trained graph classifiers. Experiments conducted on six benchmark datasets demonstrate that {\em{M-Evolve}} helps existing graph classification models alleviate over-fitting when training on small-scale benchmark datasets and 
yields an average improvement of 3$\sim$12\% accuracy on graph classification tasks.

\end{abstract}

\ccsdesc[500]{Mathematics of computing~Graph algorithms}
\ccsdesc[500]{Computing methodologies~Supervised learning by classification}

\keywords{Graph Classification; Data Augmentation; Model Evolution} 


\maketitle

\section{Introduction}
Graph classification, or network classification, has recently attracted considerable attention from different fields like bioinformatics~\cite{borgwardt2005protein} and chemoinformatics~\cite{duvenaud2015convolutional}.
For instance, in bioinformatics, proteins or enzymes can be represented as labeled graphs, in which vertices are atoms and edges represent chemical bonds that connect atoms.
The task of graph classification is to classify these molecular graphs according to their chemical properties like carcinogenicity, mutagenicity and toxicity.

However, in bioinformatics and chemoinformatics, the scale of the known benchmark graph datasets is generally in the range of tens to thousands, which is far from the scale of real-world social network datasets like COLLAB and IMDB~\cite{yanardag2015deep}.
Despite the advances of various graph classification methods, from graph kernels, graph embedding to graph neural networks, the limitation of data scale makes them easily fall into the dilemmas of over-fitting and undergeneralization.

To solve the above problem, we take an effective approach
to study data augmentation on graphs and develop two graph augmentation methods, called {\em{random mapping}} and {\em{motif-similarity mapping}}, respectively. 
The idea is to generate more virtual data for small datasets via heuristic modification of graph structures.
Since the generated graphs are artificial and treated as weakly labeled data, their availability remains to be verified.
Therefore, we introduce a concept of label reliability, which reflects the matching degree between examples and their labels, to filter fine augmented examples from generated data.
Furthermore, we introduce a model evolution framework, named {\em{M-Evolve}}, which combines graph augmentation, data filtration and model retraining to optimize classifiers. We demonstrate that {\em{M-Evolve}} achieves a significant improvement of performance on graph classification.

The main contributions of our work are summarized as follows:
\begin{itemize}[leftmargin=10pt]
  \item 
  We effectively utilize the technique of data augmentation on graph classification, and develop two methods to generate effective weakly labeled data for graph benchmark datasets.
  \item
  We propose a generic model evolution framework named {\em{M-Evolve}} for enhancing graph classification, which can be easily combined with existing graph classification models.
  \item
  We conduct experiments on six benchmark datasets.
  Experimental results demonstrate the superiority of {\em{M-Evolve}} in helping five graph classification algorithms to achieve significant improvement of performances.
\end{itemize}
\vspace{-5pt}
\section{METHODOLOGY}
\begin{figure*}[htp]
	\centering
  \includegraphics[width=\textwidth]{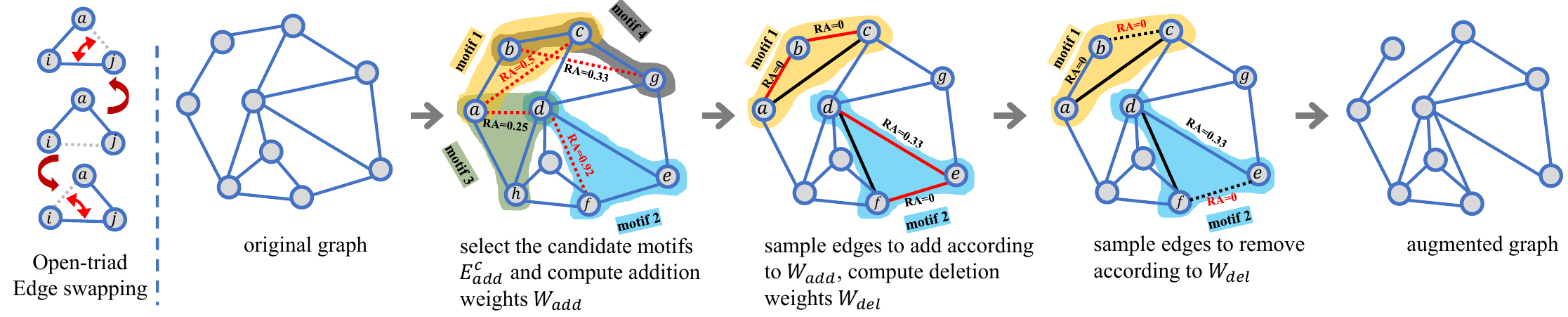}
  \setlength{\abovecaptionskip}{-7pt}
  \caption{Schematic depiction of graph augmentation. 
           {\em{Left}}: Open-triad motif and heuristic edge swapping. 
           {\em{Right}}: An example for graph augmentation via motif-similarity mapping; red lines is the candidates and black lines is the modified edges.}
  \label{fig:similarity-mapping}
\vspace{-14pt}
\end{figure*}
Let $G=(V,E)$ be an undirected and unweighted graph, which consists of a vertex set $V=\left\{v_{i} \mid i=1, \ldots, n\right\}$ and an edge set $E=\left\{e_{i} \mid i=1, \ldots, m\right\}$.
The topological structure of graph $G$ is represented by an $n \times n$ adjacency matrix $A$ with $A_{ij} = 1$ if $(i,j) \in E$ and $A_{ij}=0$ otherwise.
Dataset that contains a series of graphs is denoted as $D=\{(G_{i}, y_{i}) \mid i=1, \ldots , t\}$, where $y_i$ is the label of graph $G_i$.
For $D$, an upfront split will be applied to yield disjoint training, validation and testing set, denoted as $D_{\textit {train}}$, $D_{\textit {val}}$ and $D_{\textit {test}}$, respectively.
The original classifier $C$ will be pre-trained on $D_{\textit {train}}$ and $D_{\textit {val}}$.

\noindent \textbf{Problem Definition:} We explore data augmentation technique for graph classification problem with heuristic paradigm and consider optimizing graph classifier.
Specifically, we aim to update a classifier with augmented data, which are first generated via graph augmentation and then filtered in terms of their label reliability. 
During graph augmentation, our purpose is to map the graph $G \in D_{\textit{train}}$ to a new graph $G'$ with the formal format: $f:(G,y) \mapsto (G',y)$. 
We treat the generated graphs as weakly labeled data and classify them into two groups via a label reliability threshold $\theta$ learnt from $D_{\textit{val}}$.
Then the augmented set $D_{\textit{train}}'$ filtered from generated graph pool $D_{\textit{pool}}$ will be merged with $D_{\textit {train}}$ to produce the training set:
\begin{equation}
  \setlength{\abovedisplayskip}{3pt}
  \setlength{\belowdisplayskip}{3pt}
  D_{\textit{train}}^{\textit{new}} = D_{\textit{train}} + D_{\textit{train}}'\ , \quad
  D_{\textit{train}}' \subset D_{\textit{pool}} \ .
\end{equation}
Finally, we finetune or retrain the classifier with $D_{\textit{train}}^{\textit{new}}$, and evaluate it on the testing set $D_{\textit {test}}$.

\subsection{Graph Augmentation}
Graph augmentation aims to expand training data via artificially creating more reasonable virtual data from a limited set of graphs.
In this paper, we consider augmentation as a topological mapping, which is conducted via heuristic modification of graph structure.
In order to ensure the approximate reasonability of the generated virtual data, our graph augmentation will follow these principles:
1) edge modification, where $G'$ is a partially modified graph with some of the edges added/removed from $G$;
2) structure property preservation, where augmentation operation keeps the graph connectivity and the number of edges constant.
%
During edge modification, those edges removed from graph are sampled from the candidate edge set $E_{\textit{del}}^{\textit{c}}$, while the edges added to graph are sampled from the candidate pairwise vertices set $E_{\textit{add}}^{\textit{c}}$. The construction of candidate sets varies for different methods, as further discussed below.

\subsubsection{\rm{\textbf{Random Mapping}}}
Here, consider {\em{random mapping}} as a simple baseline. 
The candidate sets are constructed as follows:
\begin{equation}\label{random-candidate}
  \setlength{\abovedisplayskip}{3pt}
  \setlength{\belowdisplayskip}{3pt}
  E_{\textit{del}}^{\textit{c}} =  E \ , \quad
  E_{\textit{add}}^{\textit{c}} = \{(v_i, v_j) \mid A_{ij}=0;  \ i \neq j \} .
\end{equation}
Notably, in random scenario, $E_{\textit{del}}^{\textit{c}}$ is actually the edge set of graph, and $E_{\textit{add}}^{\textit{c}}$ is the set of virtual edges which consist of unlinked pairwise vertices. Then, one can get the set of edges added/removed from $G$ via sampling from the candidate sets randomly:
\begin{equation}\label{random-edgeset}
  \setlength{\abovedisplayskip}{3pt}
  \setlength{\belowdisplayskip}{3pt}
	\begin{aligned}
		&E_{\textit{del}} = \{e_i \mid i=1, \ldots, \lceil m\cdot\beta\rceil \} \subset E_{\textit{del}}^{\textit{c}} \ , \\
		&E_{\textit{add}} = \{e_i \mid i=1, \ldots, \lceil m\cdot\beta\rceil\} \subset E_{\textit{add}}^{\textit{c}} \ ,
	\end{aligned}
\end{equation}
where $\beta$ is the budget of edge modification and $\lceil x\rceil = \mathbf{ceil}(x)$.
Finally, based on the {\em{random mapping}}, the connectivity structure of the original graph is modified to generate a new graph:
\begin{equation}\label{eq:graph-update}
  \setlength{\abovedisplayskip}{3pt}
  \setlength{\belowdisplayskip}{3pt}
	G' = (V, (E \cup E_{\textit{add}}) \backslash E_{\textit{del}}) \ .
\end{equation}

\subsubsection{\rm{\textbf{Motif-Similarity Mapping}}}
Graph motifs are sub-graphs that repeat themselves in a specific graph or even among various graphs. 
Each of these sub-graphs, defined by a particular pattern of interactions between vertices, may describe a framework in which particular functions are achieved efficiently.
In this paper, we just consider open-triad motifs with chain structures. 
As shown in the left of Figure~\ref{fig:similarity-mapping}, open-triad
$\mathop{\wedge}_{ij}^{a}$ is equivalent to length-2 paths emanating from the head vertex $v_i$ that induce a triangle.

%
The {\em{motif-similarity mapping}} aims to finetune these motifs to approximately equivalent ones via edge swapping.
During edge swapping, edge addition takes effect between the head and the tail vertices of the motif, while edge deletion removes an edge in the motif via weighted random sampling.
For all open-triad motifs $\mathop{\wedge}_{ij}$ which has head vertex $v_i$ and tail vertex $v_j$, the candidate pairwise vertices set is denoted as:
\begin{equation}\label{eq:motif-mapping-cand-add}
  \setlength{\abovedisplayskip}{3pt}
  \setlength{\belowdisplayskip}{3pt}
	E_{\textit{add}}^{\textit{c}} = \{(v_i, v_j) \mid A_{ij}=0, A_{ij}^{2} \neq 0 ; i\neq j\} \ .
\end{equation}
Then one can get $E_{\textit{add}}$, the set of edges added to $G$, via weighted random sampling from $E_{\textit{add}}^{\textit{c}}$.
For each $\mathop{\wedge}_{ij}$ involving pairwise vertices $(v_i, v_j)$ in $E_{\textit{add}}$, we remove one edge from it via weighted random sampling and all of these removed edges constitute $E_{\textit{del}}$. 

Notably, we assigns all entries in $E_{\textit{add}}^{\textit{c}}$ and $\mathop{\wedge}_{ij}$ with relative sampling weights which are associated with the vertex similarity scores. 
Specifically, before sampling, we compute the similarity scores over all entries in $E_{\textit{add}}^{\textit{c}}$ using {\emph{Resource Allocation (RA)}} index which has been proven its superiority among several local similarity indices in~\cite{zhou2009predicting}.
For each entry $(v_i, v_j)$ in $E_{\textit{add}}^{\textit{c}}$, the {\emph{RA}} score $s_{ij}$ and addition weight $w_{i j}^{\textit{add}}$ can be computed as follows:
\begin{equation}\label{eq:cal-add}
  \setlength{\abovedisplayskip}{2pt}
  \setlength{\belowdisplayskip}{3pt}
	\begin{aligned}
		&s_{i j}=\sum\nolimits_{z \in \Gamma(i) \cap \Gamma(j)} \frac{1}{d_z}, \ S=\{s_{i j} \mid \forall(v_i, v_j) \in E_{\textit{add}}^{\textit{c}} \},\\
		&w_{i j}^{\textit{add}}=\frac{s_{i j}}{\sum_{s \in S} s}, \ W_{\textit{add}}=\{w_{i j}^{\textit{add}} \mid \forall(v_i, v_j) \in E_{\textit{add}}^{\textit{c}} \} ,
	\end{aligned}
\end{equation}
where $\Gamma(i)$ denotes the one-hop neighbors of $v_i$ and $d_z$ denotes the degree of vertex $z$. 
Weighted random sampling means that the probability for an entry in $E_{\textit{add}}^{\textit{c}}$ to be selected is proportional to its addition weight $w_{i j}^{\textit{add}}$.
Similarly, during edge deletion, the probability of edge sampled from $\mathop{\wedge}_{ij}$ is proportional to the deletion weight $w_{ij}^{\textit{del}}$ as follows:
\begin{equation}\label{eq:cal-delete}
  \setlength{\abovedisplayskip}{2pt}
  \setlength{\belowdisplayskip}{3pt}
		w_{ij}^{\textit{del}} = 1-\frac{s_{i j}}{\sum_{s \in S} s}, \ W_{\textit{del}}=\{w_{i j}^{\textit{del}} \mid \forall(v_i, v_j) \in \mathop{\wedge}\nolimits_{ij} \} ,
\end{equation}
which means that these edges with smaller {\emph{RA}} scores have more chance to be removed. Finally, the augmented graph can be obtained via Eq.~(\ref{eq:graph-update}).
It is worth noting that many other similarity indices such as CN and Katz~\cite{zhou2009predicting} can also be applied into this scheme.

\subsection{Model Evolution}
\subsubsection{\rm{\textbf{Data Filtration}}}
Due to the topology dependency of graph structured data, the examples generated via graph augmentation may lose original semantics. By assigning the label of the original graph to the generated graph directly during graph augmentation, one cannot determine whether the assigned label is reliable. Therefore, the concept of label reliability is employed here to measure the matching degree between examples and labels.

Each graph $G_i$ in $D_{\textit{val}}$ will be fed into classifier $C$ to obtain the prediction vector $\mathbf{p}_i \in \mathbb{R}^{|Y|}$, which represents the probability distribution how likely an input example belongs to each possible class.
$|Y|$ is the number of classes for labels. 
Then a probability confusion matrix $\mathbf{Q} \in \mathbb{R}^{|Y|\times |Y|}$, in which the entry $q_{ij}$ represents the average probability that the classifier classifies the graphs of $i$-th class into $j$-th class, is computed as follows:
\begin{equation}\label{eq:confusion-matrix}
  \setlength{\abovedisplayskip}{1.5pt}
  \setlength{\belowdisplayskip}{1.5pt}
		\mathbf{q}_k  = \frac{1}{\Omega_k} \sum_{y_i=k} \mathbf{p}_i \ , \quad
		\mathbf{Q} = [\mathbf{q}_1, \mathbf{q}_2, \ldots, \mathbf{q}_{|Y|}] \ ,
\end{equation}
where $\Omega_k$ is the number of graphs belonging to $k$-th class in $D_{\textit{val}}$ and $\mathbf{q}_k$ is the average probability distribution of $k$-th class.

The label reliability of an example $(G_i, y_i)$ is defined as the product of example probability distribution $\mathbf{p}_i$ and class probability distribution $\mathbf{q}_{y_i}$ as follows:
\begin{equation}\label{eq:label-reliability}
  \setlength{\abovedisplayskip}{1.5pt}
  \setlength{\belowdisplayskip}{1.5pt}
  r_i = {\mathbf{p}_i}^{\top} \mathbf{q}_{y_i} \ .
\end{equation}

A threshold $\theta$ used to filter the generated data is defined as:
\begin{equation}\label{eq:threshold}
  \setlength{\abovedisplayskip}{1.5pt}
  \setlength{\belowdisplayskip}{1.5pt}
  \theta = \arg \min_\theta \sum_{(G_i, y_i) \in D_{\textit{val}}} \Phi[(\theta-r_i) \cdot g(G_i, y_i)] \ ,
\end{equation} 
where $g(G_i, y_i)=1$ if $C(G_i)=y_i$ and $g(G_i, y_i)=-1$ otherwise, and $\Phi(x)=1$ if $x>0$ and $\Phi(x)=0$ otherwise.

\subsubsection{\rm{\textbf{Framework}}}
Model evolution aims to optimize classifiers via graph augmentation, data filtration and model retraining iteratively, and ultimately improve the performance on graph classification task.
The procedure of {\em{M-Evolve}} is shown in Algorithm~\ref{alg:model-evolution}.

%
\section{EVALUATION}
\subsection{Experimental Setup}
\subsubsection{\rm{\textbf{Data}}}
We evaluate our methods on six benchmark datasets: 
Mutag, PTC-MR, ENZYMES, KKI, Peking-1 and OHSU~\cite{kersting2016benchmark}.
The specifications of these datasets are given in Table~\ref{tb:dataset}, 
where $\mathit{bias}$ is the proportion of the dominant class.
\subsubsection{\rm{\textbf{Graph Classification Methods.}}}
Five graph classification methods, i.e., SF~\cite{de2018simple}, Graph2Vec~\cite{narayanan2017graph2vec}, NetLSD~\cite{tsitsulin2018netlsd}, Gl2Vec~\cite{tu2019gl2vec}, Diffpool~\cite{ying2018hierarchical}, are used in the experiments.
The first two are graph embedding, the middle two are kernel models, and the last one is GNN model.
For all graph kernel and embedding methods, we implement graph classification by using the following machine learning classifiers: SVM based on radial basis kernel (SVM), Logistic Regression (Log), $k$-Nearest Neighbors (KNN) and Random Forest (RF).
\subsubsection{\rm{\textbf{Parameter Settings.}}}
Each dataset is split into training, validation and testing sets with a proportion of 7:1:2. 
We repeat 5-fold cross validation 10 times and report the average accuracy across all trials.
For all kernel and embedding methods, the feature dimension is set to 128. 
We set the budget of edge modification $\beta$ as 0.15.
Furthermore, the evolution iterations $T$ is set to 5.
\begin{algorithm}[th]
  \caption{{\em{M-Evolve}}}  
  \LinesNumbered  
  \label{alg:model-evolution} 
  \KwIn{Training set $D_{\textit{train}}$, validation set $D_{\textit{val}}$, graph augmentation function $f$, evolution iterations $T$.}  
  \KwOut{Evolutionary model $C'$.}  	
  Pre-training classifier $C$ using $D_{\textit{train}}$ and $D_{\textit{val}}$ \;
  Initalize $\textit{iteration} = 0$\;
  \For{$\textit{iteration} < T$}
  {
    Graph augmentation: $D_\textit{pool} \leftarrow f(D_{\textit{train}})$ \;
    For all graphs $G_i$ in $D_{\textit{val}}$ classified by $C$, get $\mathbf{p}_i$ \;
    Compute probability confusion matrix $\mathbf{Q}$ via Eq.~\ref{eq:confusion-matrix} \;
    For all graphs $G_i$ in $D_{\textit{val}}$ classified by $C$, get $r_i$ via Eq.~\ref{eq:label-reliability}\;
    Compute the label reliability threshold $\theta$ via Eq.~\ref{eq:threshold} \;
    For all examples $(G_i, y_i)$ in $D_{\textit{pool}}$ classified by $C$, compute $r_i$, \textbf{if} $r_i > \theta$, $D_{\textit{train}}.append((G_i, y_i))$ \;
    Get the evolutionary classifier: $C' \leftarrow \mathsf{retrain}(C ,D_{\textit{train}})$ \;
    $\textit{iteration} \leftarrow \textit{iteration} + 1$ \;
    $C \leftarrow C'$ \;
  }
  \textbf{end} \;
  \textbf{return} $C'$;  
\end{algorithm} 
\begin{table}[htp]
  \vspace{-9pt}
  \setlength{\abovecaptionskip}{3pt}
  \setlength{\belowcaptionskip}{3pt}
  \centering
  \caption{Dataset properties.}
  \label{tb:dataset}
  \resizebox{\linewidth}{!}{%
  \begin{tabular}{c|c|ccccc} 
  \hline
  Collections                                                                     & Dataset  & \multicolumn{1}{l}{$|D|$} & \multicolumn{1}{l}{$|Y|$} & \multicolumn{1}{l}{$\textit{Avg.}|V|$} & \multicolumn{1}{l}{$\textit{Avg.}|E|$} & \multicolumn{1}{l}{\textit{bias} (\%)}  \\ 
  \hline
  \multirow{3}{*}{\begin{tabular}[c]{@{}c@{}}Chemical  ~\\Compounds\end{tabular}} & MUTAG    & 188                       & 2                         & 17.93                                  & 19.79                                  & 66.5                          \\
                                                                                  & PTC-MR   & 344                       & 2                         & 14.29                                  & 14.69                                  & 55.8                          \\
                                                                                  & ENZYMES  & 600                       & 6                         & 32.63                                  & 62.14                                  & 16.7                          \\ 
  \hline
  \multirow{3}{*}{Brain}                                                          & KKI      & 83                        & 2                         & 26.96                                  & 48.42                                  & 55.4                          \\
                                                                                  & Peking-1 & 85                        & 2                         & 39.31                                  & 77.35                                  & 57.6                          \\
                                                                                  & OHSU     & 79                        & 2                         & 82.01                                  & 199.66                                 & 55.7                          \\
  \hline
  \end{tabular}}
\end{table}
\begin{table*}
  \setlength{\abovecaptionskip}{3pt}
  \setlength{\belowcaptionskip}{3pt}
  \centering
  \caption{Graph classification results of original and evolutionary model.}
  \label{tb:result}
  \resizebox{\textwidth}{!}{%
  \begin{tabular}{c|c|cccc|cccc|cccc|cccc|c|c} 
  \hline
  \multicolumn{1}{c|}{\multirow{3}{*}{Dataset}} & \multicolumn{1}{c|}{\multirow{3}{*}{Mapping}} & \multicolumn{17}{c|}{Graph Classification Model}                                                                                                                                                         & \multirow{3}{*}{Avg
      RIMP}  \\ 
  \cline{3-19}
  \multicolumn{1}{c|}{}                         & \multicolumn{1}{c|}{}                         & \multicolumn{4}{c|}{SF} & \multicolumn{4}{c|}{NetLSD} & \multicolumn{4}{c|}{Graph2Vec} & \multicolumn{4}{c|}{Gl2Vec}               & \multicolumn{1}{c|}{\multirow{2}{*}{Diffpool}} &                               \\ 
  \cline{3-18}
  \multicolumn{1}{c|}{}                         & \multicolumn{1}{c|}{}                         & SVM          & Log          & KNN          & RF             & SVM          & Log          & KNN          & RF                 & SVM          & Log          & KNN          & RF                    & SVM          & Log          & KNN          & RF                      & \multicolumn{1}{c|}{}                          &                               \\ 
  \hline                                                                                                                                       
  \multirow{4}{*}{MUTAG}                        & original                                      &0.822         &0.824         &0.824         &0.846           &0.823         &0.829         &0.828         &0.836               &0.737         &0.820         &0.784         &0.820                  &0.746         &0.830         &0.800         &0.817                    &0.801                                                &--                               \\
                                                & random                                        &0.853         &0.844         &0.835         &0.878           &0.855         &0.851         &0.853         &0.886               &0.756         &\textbf{0.854}&0.790         &\textbf{0.844}         &0.748         &0.842         &0.820         &0.840                    &0.810                                                &2.67\%                           \\                                              
                                                & motif-similarity                              &\textbf{0.863}&\textbf{0.849}&\textbf{0.838}&\textbf{0.890}  &\textbf{0.860}&\textbf{0.864}&\textbf{0.858}&\textbf{0.892}      &\textbf{0.759}&0.849         &\textbf{0.806}&0.842                  &\textbf{0.762}&\textbf{0.848}&\textbf{0.829}&\textbf{0.846}           &\textbf{0.831}                                       &\textbf{3.60\%}                  \\ 
  \hline                                                                                                                                       
  \multirow{4}{*}{PTC-MR}                       & original                                      &0.551         &0.566         &0.577         &0.587           &0.543         &0.578         &0.548         &0.576               &0.571         &0.518         &0.509         &0.549                  &0.572         &0.538         &0.507         &0.550                    &0.609                                                &--                               \\
                                                & random                                        &0.611         &\textbf{0.595}&\textbf{0.605}&0.617           &0.579         &0.580         &0.590         &0.607               &0.580         &0.572         &0.547         &0.592                  &0.587         &0.571         &0.527         &0.594                    &\textbf{0.639}                                       &5.82\%                           \\                                               
                                                & motif-similarity                              &\textbf{0.613}&0.589         &0.601         &\textbf{0.624}  &\textbf{0.581}&\textbf{0.581}&\textbf{0.597}&\textbf{0.620}      &\textbf{0.590}&\textbf{0.579}&\textbf{0.553}&\textbf{0.593}         &\textbf{0.587}&\textbf{0.579}&\textbf{0.545}&\textbf{0.602}           &0.630                                                &\textbf{6.60\%}                  \\ 
  \hline                                                                                                                                       
  \multirow{4}{*}{ENZYMES}                      & original                                      &0.309         &0.393         &0.287         &0.397           &0.337         &\textbf{0.248}&0.304         &0.349               &0.361         &0.253         &0.283         &0.337                  &\textbf{0.348}&0.268         &0.238         &0.318                    &0.487                                                &--                               \\
                                                & random                                        &0.347         &0.412         &0.302         &0.412           &0.351         &0.237         &0.327         &0.369               &0.336         &0.269         &\textbf{0.290}&0.346                  &0.286         &0.273         &0.259         &0.334                    &0.500                                                &2.60\%                           \\                                             
                                                & motif-similarity                              &\textbf{0.363}&\textbf{0.414}&\textbf{0.317}&\textbf{0.414}  &\textbf{0.375}&0.248         &\textbf{0.334}&\textbf{0.376}      &\textbf{0.352}&\textbf{0.270}&0.289         &\textbf{0.352}         &0.291         &\textbf{0.280}&\textbf{0.260}&\textbf{0.339}           &\textbf{0.506}                                       &\textbf{5.00\%}                  \\ 
  \hline                                                                                                                                       
  \multirow{4}{*}{KKI}                          & original                                      &0.550         &0.500         &0.520         &0.517           &0.548         &0.524         &0.512         &0.496               &0.549         &0.527         &0.524         &0.552                  &0.538         &0.502         &0.526         &0.502                    &0.523                                                &--                               \\
                                                & random                                        &\textbf{0.606}&0.544         &0.554         &0.622           &0.599         &0.535         &0.553         &0.562               &0.580         &0.568         &0.594         &0.574                  &0.556         &0.508         &0.544         &0.544                    &0.586                                                &8.10\%                           \\                                               
                                                & motif-similarity                              &0.605         &\textbf{0.559}&\textbf{0.561}&\textbf{0.649}  &\textbf{0.618}&\textbf{0.550}&\textbf{0.558}&\textbf{0.582}      &\textbf{0.587}&\textbf{0.590}&\textbf{0.603}&\textbf{0.632}         &\textbf{0.574}&\textbf{0.524}&\textbf{0.597}&\textbf{0.582}           &\textbf{0.612}                                       &\textbf{12.07\%}                 \\ 
  \hline                                                                                                                                       
  \multirow{4}{*}{Peking-1}                     & original                                      &0.578         &0.548         &0.541         &0.558           &0.605         &0.612         &0.589         &0.591               &0.572         &0.522         &0.474         &0.522                  &0.555         &0.522         &0.521         &0.521                    &0.586                                                &--                               \\
                                                & random                                        &0.660         &\textbf{0.592}&0.603         &0.627           &0.652         &\textbf{0.631}&0.662         &0.654               &0.579         &0.536         &\textbf{0.546}&0.576                  &0.584         &0.555         &\textbf{0.569}&0.607                    &\textbf{0.631}                                       &9.08\%                           \\                                              
                                                & motif-similarity                              &\textbf{0.670}&0.583         &\textbf{0.624}&\textbf{0.663}  &\textbf{0.694}&0.627         &\textbf{0.671}&\textbf{0.699}      &\textbf{0.581}&\textbf{0.565}&0.539         &\textbf{0.630}         &\textbf{0.607}&\textbf{0.563}&0.562         &\textbf{0.625}           &0.626                                                &\textbf{11.88\%}                 \\ 
  \hline                                                                                                                                       
  \multirow{4}{*}{OHSU}                         & original                                      &0.610         &0.595         &0.610         &0.667           &0.547         &0.489         &0.549         &0.581               &\textbf{0.557}&0.577         &0.585         &0.567                  &0.557         &0.541         &0.544         &0.557                    &0.543                                                &--                               \\
                                                & random                                        &\textbf{0.663}&0.635         &\textbf{0.644}&0.687           &0.575         &0.494         &0.582         &\textbf{0.641}      &0.557         &0.620         &0.602         &0.645                  &0.564         &0.595         &0.625         &0.610                    &0.600                                                &6.87\%                           \\                                               
                                                & motif-similarity                              &0.656         &\textbf{0.640}&0.636         &\textbf{0.707}  &\textbf{0.638}&\textbf{0.501}&\textbf{0.587}&0.638               &0.557         &\textbf{0.625}&\textbf{0.633}&\textbf{0.650}         &\textbf{0.572}&\textbf{0.605}&\textbf{0.625}&\textbf{0.652}           &\textbf{0.604}                                       &\textbf{8.83\%}                  \\
  \hline
\end{tabular}}
\vspace{-9pt}
\end{table*}
\subsection{Enhancement for Graph Classification}
Table~\ref{tb:result} reports the results of performance comparison between evolutionary models and original models, from which we can see that there is a significant boost in classification performance across all six datasets. 
Overall, these models combined with the proposed {\em{M-Evolve}} framework obtain higher average classification accuracy in most cases and the {\em{M-Evolve}} achieves a 97.06\% success rate on the enhancement of graph classification. 
Moreover, the far-right column gives the average relative improvement rate (Avg RIMP) in accuracy, from which we can see that the {\em{M-Evolve}} combined with {\em{motif-similarity mapping}} obtains the best results overall. As a reasonable explanation, similarity mechanism tends to link vertices with higher similarity and is capable of optimizing topological structure legitimately, while the motif mechanism achieves edge modification via local edge swapping, which has less effect on the degree distribution of the graph.
\begin{figure}[htp]
	\centering
  \includegraphics[width=\linewidth]{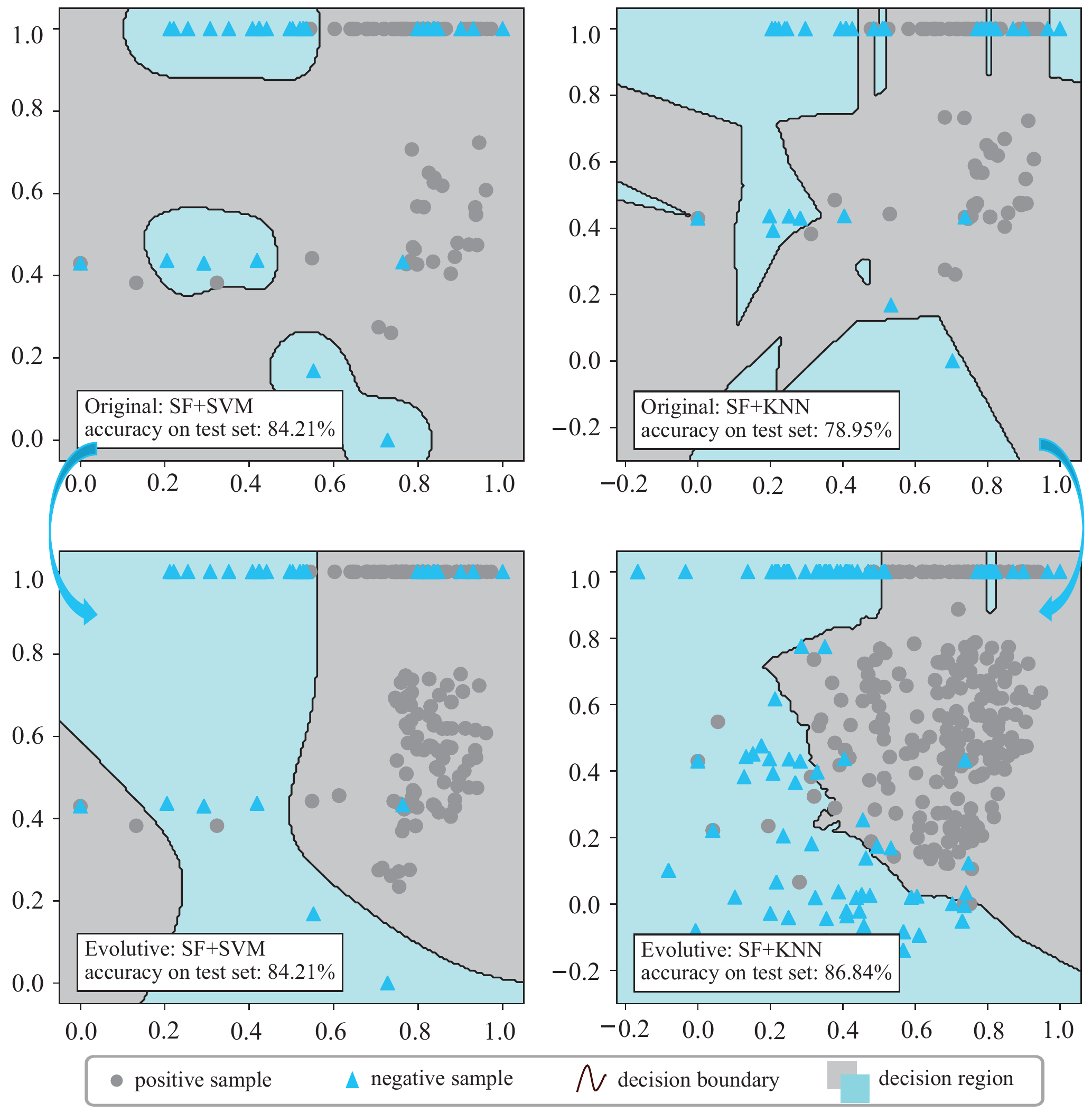}
  \setlength{\abovecaptionskip}{-7pt}
  \caption{Visualization of training data distribution and decision boundaries of graph classifiers on MUTAG dataset.}
  \label{fig:dr}
\vspace{-15pt}
\end{figure}

Furthermore, we visualize the data distribution and the decision boundary of models before and after model evolution, as shown in Figure~\ref{fig:dr}, to investigate how the {\em{M-Evolve}} framework achieves interpretable enhancement of graph classification. 
Due to space limit, we only present the visualization results of graph classification based on these combinations (\textit{SF} + \{\textit{SVM}, \textit{KNN}\}).
As we can see, there is a significant increase in the scale of dataset, indicating that graph augmentation effectively enriches the training data and the new data distribution is more conducive to the training of classifiers. Moreover, the decision regions of the non-dominant class are fragmented and scattered in the original models. During model evolution, scattered regions tend to merge, and the original decision boundaries are optimized to smoother ones.

In summary, graph augmentation can efficiently increase
the data scale, indicating its ability in enriching data distribution. And the entire {\em{M-Evolve}} framework is capable of optimizing the decision boundaries of the classifiers and ultimately improving their generalization performances.

\section{CONCLUSION}
In this paper, we introduce a concept of graph augmentation in graph structured data and present two heuristic algorithms to generate weakly labeled data for small-scale benchmark datasets via heuristic transformation of graph structure. 
Furthermore, we propose a generic model evolution framework that combines graph augmentation, data filtration and model retraining to optimize pre-trained graph classifiers. Experiments conducted on six benchmark datasets demonstrate that our proposed framework behaves surprisingly well and helps existing graph classification models alleviate over-fitting when training on small-scale benchmark datasets and achieve significant improvement of classification performance. 
\begin{acks}
  This work was partially supported by the National Natural Science Foundation of China under Grant 61973273, 
  and by the Zhejiang Provincial Natural Science Foundation of China under Grant LR19F030001.
\end{acks}

\bibliographystyle{ACM-Reference-Format}
\balance
\bibliography{sample-base}










\end{document}